\documentclass[aps, prl, showpacs, nofootinbib, preprintnumbers,twocolumn]{revtex4}
\usepackage{graphicx,color}
\usepackage{amsmath,amssymb}
\newcommand{\VEV}[1]{\langle #1 \rangle}
\newcommand{\tr}{\mathrm{tr}}
\newcommand{\Tr}{\mathrm{Tr}}
\newcommand{\diag}{\mathrm{diag}}
\newcommand{\gBL}{g_{B-L\rule{0mm}{7pt}}}
\newcommand{\gBLsq}{g^2_{B-L\rule{0mm}{6pt}}}

\newcommand{\gtld}{g_{\mathrm{mix}}}

\newcommand{\be}{\begin{eqnarray}}
\newcommand{\ee}{\end{eqnarray}}
\newcommand{\n}{\nonumber \\}
\begin{document}
\title{Radiative symmetry breaking at the Fermi scale \\
and \\ flat potential at the Planck scale} 
\preprint{KEK-TH1678}
\preprint{OU-HET795}
\author{Michio Hashimoto${}^1$}
\author{Satoshi Iso${}^2$}
\author{Yuta Orikasa${}^3$}
\affiliation{${}^1$
Chubu University,  
Kasugai-shi,  Aichi, 487-8501, JAPAN }

\affiliation{${}^2$
Theory Center, KEK and
Sokendai,  
Tsukuba, Ibaraki 305-0801, JAPAN }

\affiliation{${}^3$
Department of Physics, 
Osaka University,  Toyonaka, Osaka 560-0043, JAPAN}

\date{\today}

\begin{abstract}
We investigate a possibility of the ``flatland scenario'',
in which the electroweak gauge symmetry is radiatively broken via
the Coleman-Weinberg mechanism starting from a completely flat Higgs
potential at the Planck scale.
We show that the flatland scenario is realizable only when 
 an inequality $K<1$ among the coefficients of 
the $\beta$-functions is satisfied.
We show several models satisfying the condition.
\end{abstract}
\maketitle

 {\it Introduction.---}
The Brout-Englert-Higgs (BEH) mechanism in the context of 
the Standard Model (SM) is the source of the electroweak symmetry breaking 
(EWSB) and predicts appearance of the Higgs boson~\cite{BEH-mechanism}.
Recently the LHC experiment announced the discovery of a new particle 
like the Higgs boson in the SM and 
its mass is now determined to be $m_H = 125$--$126$~GeV~\cite{Higgs-discover}.
This mass causes the so-called stability problem of the SM vacuum.
Compared with the value of the condensation $v=246$ GeV, the mass is relatively
small and the Higgs potential seems to be shallow. 
The shallowness indicates instability of the potential against the radiative corrections, and
indeed, if we calculate the running quartic coupling of the Higgs boson $\lambda_H(\mu)$
where $\mu$ is the renormalization scale, 
it tends to vanish at a very high energy scale \cite{stability}.  
Within the uncertainties of the top quark mass and the strong gauge coupling,
the running Higgs coupling seems to vanish asymptotically 
 $\lambda_H(\mu) \rightarrow 0$
near the Planck scale $\mu \rightarrow M_{Pl}$. 

Another important hint for the origin of the  Higgs potential comes from the naturalness problem.
The Higgs mass receives large radiative corrections by, if exist, heavy particles  
coupled to the Higgs boson. The supersymmetry in the TeV scale gives 
a beautiful solution to the naturalness problem,
but the LHC and other precision experiments have strong constraints on their masses. 
Also, the Technicolor scenario is faced with the difficulties of 
the $S$-parameter and the smallness of the Higgs mass.
Recently  alternative solutions to the naturalness problem are widely discussed \cite{naturalness}. 
Suppose that the UV completion theory (which may be beyond the ordinary
field theories like the string theory) is connected with the SM sector
in a way that the SM has no dimensionful parameters. 
Then if no large intermediate mass scales exist between the SM and the UV completion theory,
no large logarithmic corrections violating the multiplicative renormalization of the 
Higgs mass term are generated and
the SM becomes free from the naturalness problem.
Such a model based on the idea
is called a classically conformal model with no intermediate scales \cite{ccm,Iso:2009ss}.

Motivated by the stability of the vacuum and the naturalness problem, we explore 
a possibility \cite{Iso:2012jn,Chun:2013soa}
that the EW symmetry is radiatively broken in the infrared (IR) region 
via the Coleman-Weinberg mechanism (CWM)~\cite{CW} 
starting from a  flat scalar potential 
in the ultraviolet (UV) region.
We call it a "flatland scenario". 
It is nontrivial to construct such a model  because 
the scalar quartic coupling must be tuned to become very small
both in the IR and the UV regions.
It is well known that the CWM does not work within the SM
because of the large top Yukawa coupling.
Thus we need to extend the SM by introducing an additional
sector in which the dynamical mass generation occurs.
In this letter, we show that a certain inequality $K<1$ must be satisfied
among the coefficients of the $\beta$ functions 
in order to realize the flatland scenario.

{\it A necessary condition for the flatland.---}
We first consider a system of a complex scalar field $\Phi$ charged under the Abelian gauge field.
We further introduce a charged fermion $\psi$ with a (Majorana) Yukawa coupling $y$ to the scalar field.
The gauge coupling is denoted by $g$ and $\lambda$ is the quartic coupling of the scalar field.
The RGE's can be written in terms of model-dependent positive constants
$a,b,c,d,f$ as
\be
  \beta_g &\equiv& \mu \frac{\partial }{\partial \mu} g =
  \frac{a}{16 \pi^2}  g^3, 
  \label{RGE1} \\
  \beta_y &\equiv& \mu \frac{\partial }{\partial \mu} y =
  \frac{y}{16 \pi^2} \bigg[\, b y^2 - c g^2\,\bigg],  \label{RGE2} \\
  \beta_\lambda &\equiv& \mu \frac{\partial }{\partial \mu} \lambda =
  \frac{1}{16 \pi^2} \bigg[\, - d y^4 + f g^4 + \cdots \,\bigg]\,.
   \label{RGE3}
\ee
The dots in $\beta_\lambda$ include terms 
$\lambda^2$, $\lambda g^2$, etc. They are irrelevant in 
the following discussions because the quartic coupling $\lambda$ becomes very small
both in the IR region where the CWM occurs and 
in the UV region where we impose $\lambda(\Lambda)=0$ at a UV scale $\Lambda$.
Introducing  $t=(\ln \mu/M)/16 \pi^2$ ($M$ is an IR scale)
and the new variable $r=y/g$, we can rewrite eqs. (\ref{RGE2}),(\ref{RGE3})  into 
\be
\dot{r}=b r g^2 (r^2 -r_c^2),  \ \ \dot{\lambda} = d g^4 (r_0^4-r^4), 
\label{RGEmod}
\ee
where the dot denotes  $t$-derivative and $r_c, r_0$ are the zeros of 
the $\beta$-functions of $r=y/g$ and $\lambda$ respectively; 
\be
 r_c = \sqrt{\frac{a+c}{b}}, \ \ r_0=\left( \frac{f}{d} \right)^{1/4}.
\ee
From the first equation of (\ref{RGEmod}), we can see that $r=r_c$ is an IR fixed point
\cite{Pendleton:1980as} for the ratio $r$, i.e., if $r>r_c$ (or $r<r_c)$, $r(t)$ is an increasing 
(decreasing) function of $t$.

\begin{figure}[t]
  \begin{center}
  \resizebox{0.3\textheight}{!}
            {\includegraphics{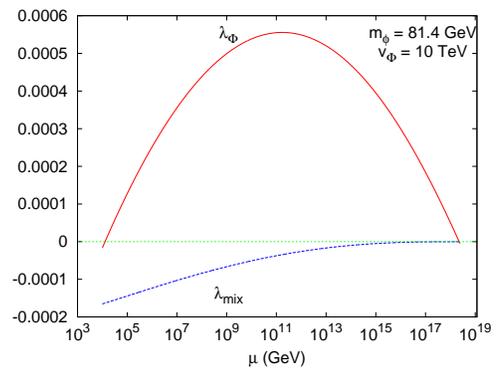}}
  \end{center}
  \caption{The RG flows of $\lambda_\Phi$ and $\lambda_{\mathrm{mix}}$ 
  in  $(2,1,1)$ model. $v_\Phi=10$~TeV, $m_\phi=81.4$~GeV and $y_M(v_\Phi)=0.378$.
  $K$ is close to $1$ and the  quartic coupling $\lambda_\Phi$ is small.
  Accordingly, the scalar mass becomes very light compared with $v_\Phi$.
  \label{flow-3rd-2nd-lam2-3}}
\end{figure}

Now let us study the condition for the flatland scenario,
namely the condition that the gauge symmetry is spontaneously broken with a
vacuum expectation value $\VEV{\Phi} = M/\sqrt{2}$ via the CWM  in a model with a vanishing quartic coupling 
at the UV scale, $\lambda(t_{UV})=0$.  
A typical behavior of the running quartic coupling $\lambda_\Phi$
is given in figure \ref{flow-3rd-2nd-lam2-3}. 
This figure shows that 
in the IR region $t \sim 0$, the $\beta$-function of $\lambda$ must 
be positive $\beta_\lambda(t=0) >0$ while being negative $\beta_\lambda(t_{UV})<0$
in the UV. Hence $r(t)$ must satisfy the inequalities $ r(t=0)<r_0 < r(t_{UV}). $
Then, since $r(t)$ is an increasing function of $t$, it must be larger than $r_c$;
\be
r_c<  r(t=0)<r_0 < r(t_{UV}). 
\ee
Hence we obtain a necessary condition among the coefficients of the 
$\beta$-functions  for the flatland scenario;
\be
K =\left( \frac{r_c}{r_0} \right)^2 =\frac{a+c}{b} \sqrt{\frac{d}{f}} <1.
\label{nec-cond}
\ee
Unless the inequality is satisfied, the radiative symmetry breaking, namely the CWM, does not
occur starting from the flat potential at the UV scale. 

It is furthermore required that the ratio $r(0)=y(0)/g(0)$ at the IR scale is tuned 
to lie in-between $r_c$ and $r_0$.
This fact is followed by the smallness of the scalar boson mass 
in a model with  $K$ close to 1, $K \lesssim 1.$
In the CWM,  the scalar boson acquires its mass 
proportional to the $\beta$ function
(see, e.g., eq.(4) in the first paper of \cite{Iso:2009ss}\footnote{The $\beta$-function
is given by $\beta (0)=8B$ in the notation.}),
\be
m_\phi^2 =  \beta_\lambda(0) M^2 
=\frac{  M^2}{16 \pi^2} \left( -d y^4(M)+f g^4(M) \right) >0.
\label{mphi-mzp-mnu1}
\ee
Hence, if $K \lesssim 1$, $r_c \sim r_0$ and $r(0)$ is  also
very close to $r_0$. Thus $\beta_\lambda(0) \sim 0$ and 
the scalar mass of eq.(\ref{mphi-mzp-mnu1})
becomes tiny. We will see such a situation explicitly below.

Let us now evaluate the ratio $K$ for 
the $B-L$ (baryon minus lepton number) 
models with the Majorana Yukawa interactions\footnote{
$Y_M^{ij}$ defined in the present letter is equal to 
$Y_N^{ij}/2$ in Ref.\cite{Iso:2012jn}.}
between the SM singlet complex scalar field $\Phi$ and 
the right-handed neutrinos $\nu_{R i}$,
\begin{equation}
  {\cal L}_M = - Y_M^{ij} \overline{\nu_{Ri}^c} \, \nu_{Rj} \Phi 
  + \mbox{(h.c.)} \, .
  \label{yM}
\end{equation}
The coefficients\footnote{
The coefficient $d$ of the $y_M^4$ term in $\beta_\lambda$ 
in Ref.~\cite{Basso:2010jm} was  
{\it 16 times smaller} than ours. 
We thank L. Basso for his e-mail correspondence and acknowledging
the above error  in \cite{Basso:2010jm}.
Also there is a typo in \cite{Iso:2012jn}. The coefficient of $\Tr [Y_N^4]$ in eq.(35)
is $-1$ instead of $-1/2$.
\label{error}}
 of the RGE's 
read~\cite{Iso:2009ss,Basso:2010jm,Iso:2012jn}
\be
  a &=& \frac{32}{9} N_g + \frac{4}{3} N_\Phi, \quad  b = 4+2N_\nu, \n  
  c &=& 6, \quad d=16N_\nu, \quad f=96,
  \label{abcdf}
\ee
where $N_g$ is the number of generations coupled to
the $U(1)_{B-L}$ gauge field and $N_\Phi \; (=1)$ is the number of $\Phi$.
$N_\nu$  stands for the number of the right handed neutrinos 
having relevant Majorana couplings. For simplicity, 
we take $Y_M^{ij} = \diag (y_M,\cdots,y_M, 0, \cdots, 0)$ and
$\tr [(Y_M^{ij})^{2}] = N_\nu y_M^{2}$ etc.
We denote these models by $(N_g, N_\Phi, N_\nu)$.

The ratios $K$ for various $B-L$ models are listed in the Table \ref{tab1}.
In the $(3,1, N_\nu)$  models, 
K is always larger than $1$.
Hence the flatland scenario does not occur 
\footnote{The numerical result  in  \cite{Chun:2013soa} showed 
the flatland scenario in the $(3,1, N_\nu)$  model, but it is because
the wrong coefficient of the $\beta$-function  in \cite{Basso:2010jm} was used there.
If $d$ were 16 times smaller, it would give $K<1$. 
See also the corrigendum to \cite{Chun:2013soa}.
\label{Chun} } 
in the models with $N_g=3$.

In order to satisfy the condition (\ref{nec-cond}), we will consider
two possibilities, (1) decreasing $a$, or (2) increasing $b$ (or $N_\nu$).
The first possibility is realized for smaller $N_g$.
From the Table \ref{tab1}, we see that  $K < 1$ for $N_g=1,2$ and $N_\nu=1$.
These are the candidates of the flatland scenario.
The second possibility is to introduce SM and $B-L$ singlet fermions 
so that  $N_\nu$ is larger than three.

\begin{table}
$$
  \begin{array}{c|l} \hline
    N_g = 3 &
    \begin{array}{c|l}
      N_\nu = 1 & K = \sqrt{3/2} \simeq 1.22 \\
      N_\nu = 2 & K = 3\sqrt{3}/4 \simeq 1.30 \\
      N_\nu = 3 & K = 9\sqrt{2}/10 \simeq 1.27 \\
    \end{array} \\ \hline
    N_g = 2 &
    \begin{array}{c|l}
      N_\nu = 1 & K = 65\sqrt{6}/162 \simeq 0.98 \\
      N_\nu = 2 & K = 65\sqrt{3}/108 \simeq 1.04 \\
    \end{array} \\ \hline
    N_g = 1 &
    \begin{array}{c|l}
      N_\nu = 1 & K = 49\sqrt{6}/162 \simeq 0.74 \\
    \end{array} \\ \hline
  \end{array}
$$
  \caption{Values of $K$ for various $B-L$ models $(N_g, N_\Phi, N_\nu)$. 
   We fixed to $N_\Phi=1$, which minimizes $K$ with respect to $N_\Phi$.
   The flatland  scenario is possible only when $K<1$.}
  \label{tab1}
\end{table}

{\it Flatland scenario.---}
We investigate more detailed analysis of the RGE's in the flatland scenario  and
show how the EWSB is triggered by the $B-L$ symmetry breaking.
The Lagrangian of the $B-L$ model is 
\begin{equation}
  {\cal L} = {\cal L}_{\rm SM} + {\cal L}_{\rm kin} 
  + {\cal L}_M - V,
  \label{b-l-3gen}
\end{equation}
where ${\cal L}_{\rm SM}$ represents the SM part, 
${\cal L}_{\rm kin}$ is the kinetic terms of $\nu_R$, $\Phi$ and
the $B-L$ gauge field $B'_\mu$.
We fix $N_\Phi=1$.
The potential for the Higgs doublet $H$ and the SM singlet $\Phi$ 
is given by $V$,
\begin{equation}
V = \lambda_H |H|^4 +\lambda_\Phi |\Phi|^4
  + \lambda_{\mathrm{mix}} |H|^2 |\Phi|^2 .
 \label{V}
\end{equation}
We assumed the classical conformality, i.e.,
the mass squared terms are absent.
In the basis where the two $U(1)$ gauge kinetic terms are diagonal,
the covariant derivative is written as 
\begin{eqnarray}
  D_\mu &=& \partial_\mu - ig_3 \frac{\lambda^a}{2} G_\mu^a
  - ig_2 \frac{\tau^i}{2} W_\mu^i \nonumber \\ &&
  - i Y (g_Y B_\mu + \gtld B'_\mu)
  - i \gBL Y_{B-L} B'_{\mu}, 
\end{eqnarray}
where $Y$ and $Y_{B-L}$ denote the hypercharge and the $B-L$ charge, 
respectively.
The gauge couplings of $SU(3)_c$, $SU(2)_W$, $U(1)_Y$ and $U(1)_{B-L}$ 
are $g_3$, $g_2$, $g_Y$ and $\gBL$, respectively.
In general, the $U(1)$ gauge mixing $\gtld$ appears
owing to the loop corrections of the fermions having both charges of
$U(1)_Y$ and $U(1)_{B-L}$, even if we impose $\gtld=0$ at some scale.

For appropriate parameters investigated below, 
the CWM occurs in  the $B-L$ sector of $\Phi$~\cite{Iso:2009ss,Iso:2012jn}.
Then the EWSB takes place if 
the $H$--$\Phi$ mixing term $\lambda_{\mathrm{mix}} |H|^2 |\Phi|^2$
is negative;
\begin{equation}
  v_H^2 = \frac{- \lambda_{\mathrm{mix}}}{2\lambda_H} \, v_\Phi^2 ,
  \label{vH}
\end{equation}
where $\VEV{H} = (0, \; v_H/\sqrt{2})^T$ and
$\VEV{\Phi} = v_\Phi/\sqrt{2}$.
The Higgs mass $m_h$ is approximately given 
by $m_h^2 = 2\lambda_H v_H^2$, because the mixing 
between $H$ and $\Phi$ is tiny.
In \cite{Iso:2012jn}, we have shown that such a small and 
negative scalar mixing is radiatively generated through 
the gauge kinetic mixing of $U(1)_{B-L}$ and $U(1)_Y$.

In order to realize the flatland scenario, 
we impose vanishing of the scalar potential at a UV scale $\Lambda$,
\begin{equation}
 \lambda_H (\Lambda)=0, \quad \lambda_\Phi (\Lambda)=0, \quad
 \lambda_{\mathrm{mix}} (\Lambda)=0.
 \label{flatboundary}
\end{equation}
We also set $ \gtld (\Lambda)=0$ 
by constructing a model with no $U(1)$ kinetic mixing 
at the high energy scale $\Lambda$. 
The gauge mixing 
$\gtld \ne 0$ between $Z$ and $Z'$  is generated in the EW scale
through the RG effects. It is potentially dangerous, but
we find that the deviation of the $\rho$ parameter from unity is tiny,
at most $\delta \rho_0 \sim {\cal O}(10^{-5})$ \cite{HIO}.
The remaining  parameters are $\gBL$ and $y_M$,
which correspond to the masses of $Z'$ and $\nu_R$,
\begin{equation}
  M_{Z'} \simeq 2 \gBL v_\Phi, \quad
  M_{\nu_R} \simeq \sqrt{2}\, y_M v_\Phi .
\end{equation}

Starting from the flat potential (\ref{flatboundary}) at the UV scale,
 running couplings are obtained dynamically by solving the RGE's, 
and the value of $v_H$ in Eq.~(\ref{vH}) is predicted in terms of them.
The RG flows are controlled by the gauge coupling $g_{B-L}$
and the Majorana Yukawa coupling $y_M$ at $\Lambda$ besides the SM
parameters.
Given these two parameters at $\Lambda$, the symmetry breaking scales of 
$\Phi$ and $H$ are determined.
In order to set $v_H =  246$~GeV, we must adjust one of the two parameters
in accordance with the other. 
Hence there is only {\it one free parameter} in the model.
In particular, the CW relation
\begin{equation}
   \frac{m_\phi^2}{M_{Z'}^2}
 + N_\nu \frac{\gBLsq}{\pi^2}\frac{M_{\nu_R}^4}{M_{Z'}^4}
 \simeq \frac{3\gBLsq}{2\pi^2} 
 \label{mzp-mphi-mnuR}
\end{equation}
must hold  where we rewrote the equation (\ref{mphi-mzp-mnu1})  in terms of the physical quantities.

{\it Numerical analysis.---}
We numerically solve the RGE's of the 
$(2, 1,1)$ and $(1,1,1)$ models. 
In the following analysis, we take the UV scale at
$\Lambda=1/\sqrt{8\pi G} = 2.435 \times 10^{18}$~GeV.
Also, we fix the Higgs mass, $m_h=126.8$~GeV
and the $\overline{MS}$ mass of the top quark\footnote{
This is consistent with the indirect prediction, 
$\overline{m_t}=167.5^{+8.9}_{-7.3}$~GeV~\cite{pdg}, while
the converted value to the pole mass~\cite{Melnikov:2000qh} 
is rather small, compared to
the directly obtained value at the Tevatron/LHC.} 
$\overline{m_t}=160.4$~GeV so as to realize $\lambda_H(\Lambda)=0$
\footnote{The discrepancy between the condition $\lambda_H(\Lambda) =0$ 
and the current experimental data can be improved by considering 2-loop corrections.}.

\begin{figure}[t]
  \begin{center}
  \resizebox{0.3\textheight}{!}
            {\includegraphics{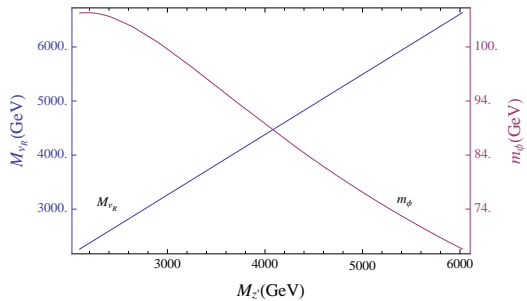}}
  \end{center}
  \caption{$M_{Z'}$ vs $M_{\nu_R}$and $m_\phi$ for the $(2,1,1)$ model.
  $M_{\nu_R}$ is almost proportional to $M_{Z'}$.
    \label{2mzpmn}}
\end{figure}

We first investigate the $(2,1,1)$ model. 
Since $K=0.98 $, the necessary condition for the flatland scenario is
barely satisfied, suggesting that the SM singlet scalar mass $m_\phi$ is
very light compared to the gauge boson mass $M_{Z'}$. 
If we take, for example, $v_\Phi=10$~TeV, 
the other parameters are numerically determined to be 
$m_\phi=81.4$~GeV, $y_M(v_\Phi)=0.378$ and $M_{Z'}=4.87$~TeV.
The scalar mass is actually light.
The RG flows of $\lambda_\Phi$ and $\lambda_{\mathrm{mix}}$ are depicted
in Fig.~\ref{flow-3rd-2nd-lam2-3}. The very small quartic coupling $\lambda_\Phi$ 
corresponds to the lightness of $m_\phi$.  
Fig.~\ref{2mzpmn} shows relations between $m_\phi, M_{\nu_R}$ and $M_{Z'}$.
When $m_\phi^2 \ll \gBLsq M_{Z'}^2$, we get approximately
$M_{\nu_R} \sim \sqrt[4]{3/(2 N_\nu)} \, M_{Z'}$ from the CW relation (\ref{mzp-mphi-mnuR}).
We also comment that $K=1$ is possible if we choose appropriately 
the second largest $Y_M^{ij}$.
Then the scalar potential becomes flat in all energy scales, and $\Phi$
becomes massless.

\begin{figure}[t]
  \begin{center}
  \resizebox{0.3\textheight}{!}
            {\includegraphics{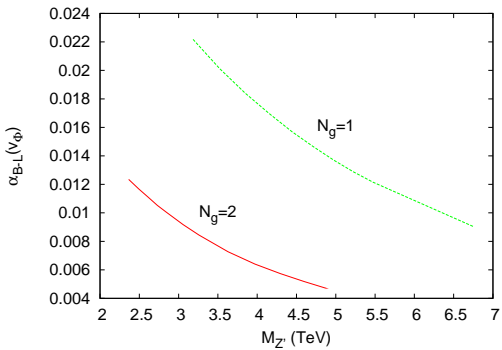}}
  \end{center}
  \caption{$M_{Z'}$ vs $\alpha_{B-L}(v_\Phi)$ for the $(2,1,1)$ and
    $(1,1,1)$ models.
  \label{2mzpalpha}}
\end{figure}

For the $(1,1,1)$ model, we obtain similar results, but not repeated here.
Figures~\ref{2mzpalpha} shows the relation between 
$M_{Z'}$ and $\alpha_{B-L}$ for the $(2,1,1)$ and $(1,1,1)$ models.
The gauge coupling is relatively larger for a fixed $M_{Z'}$ than the prediction of $(3,1,1)$ model studied in \cite{Iso:2012jn}. 

In the $(2,1,1)$ and $(1,1,1)$ models, 
the off diagonal terms in the SM yukawa couplings are not 
written in dimension four operators.
For this purpose, we may introduce an additional scalar field $S$ 
with a fractional $B-L$ charge such as 
\begin{equation}
  {\cal L}_{\rm off} =
  - Y_\nu^{(13)} \overline{\ell^{(1)}_L} \tilde{H} \nu_R^{(3)}
  \left(\frac{S}{\Lambda}\right)^k + \cdots \, .
  \label{FN}
\end{equation}
A simple model contains a scalar $S$ with $Q=1/3$ $B-L$ charge.
In this case $K$ becomes $K=0.985$, and the flatland scenario is still possible.
More generally, including $n$ additional scalars with $B-L$ charge $Q$  
into the $(2,1,1)$ model,
the ratio $K$ is modified to be $K=\sqrt{6}/108(130/3+n Q^2)$. 

The introduction of the scale $\Lambda$ in (\ref{FN}) is not favourable from the
standpoint of the classical conformality.  It can be evaded if we
regard the $N_g=1,2$ sector as extra (or ``hidden'') generations
like in Ref.~\cite{Lee:2012xn}. 
In this case, the higher-dimensional terms are not required and the model
can be compatible with the classical conformality of the SM sector. 

Another possibility is to introduce extra Higgs doublets with 
the $B-L$ charges.
Furthermore, $U(1)'$ models are not restricted only to 
the $B-L$ one~\cite{Langacker:2008yv}. 
An example of such generalisations was 
investigated in the corrigendum to \cite{Chun:2013soa}. 

The second type of possibility of the flatland scenario is to introduce $N_\nu$
singlet fermions $\psi_{sj}$  with a coupling 
$\overline{\nu^c_{Ri}} \psi_{sj} \Phi$, 
where the $B-L$ charge of $\Phi$ should be changed appropriately,
and a Majorana mass term for them
instead of the Majorana Yukawa couplings.
In such a model, the condition $K<1$ is satisfied for 
$N_\nu \sim {\cal O}(100)$.

{\it Summary.---}
The origin of the Higgs potential is one of the unsolved issues in particle physics.
We explore possibilities that the EWSB occurs starting from a completely flat potential
at a UV energy scale. The scenario, which we call the flatland scenario, is possible
only when the system satisfies an inequality
$K<1$ of eq.(\ref{nec-cond}). This is the main result of the paper. 
The condition gives a strong constraint when we construct a model of
the radiative EWSB
with a flat potential at the Planck scale \cite{flat}. 
We schematically showed some models satisfying the condition. 
More detailed analysis are investigated in a separate paper \cite{HIO}.


\begin{thebibliography}{99}
\bibitem{BEH-mechanism}
  F.~Englert and R.~Brout,
  Phys. Rev. Lett.  {\bf 13}, 321 (1964);
  P.~W.~Higgs,
  {\it ibid.} {\bf 13}, 508 (1964).

\bibitem{Higgs-discover}
  G.~Aad {\it et al.}  [ATLAS Collaboration],
  Phys. Lett. B {\bf 716}, 1 (2012);
  S.~Chatrchyan {\it et al.}  [CMS Collaboration],
  {\it ibid.} 
  B {\bf 716}, 30 (2012);
   The CMS Collaboration, HIG-13-001;
   The ATLAS Collaboration, ATLAS-CONF-2013-014.

\bibitem{stability}
 J.~Elias-Miro {\it et al.}, 
  Phys. Lett. B {\bf 709}, 222 (2012); 
C.~P.~Burgess, V.~Di Clemente and J.~R.~Espinosa,
  JHEP {\bf 0201}, 041 (2002);
  D.~Buttazzo {\it et al.}, 
  arXiv:1307.3536 [hep-ph].

\bibitem{naturalness}
 W.~A.~Bardeen,
  FERMILAB-CONF-95-391-T;
    R.~Foot, et.al. 
  Phys. Rev. D {\bf 77}, 035006 (2008);
   M.~Shaposhnikov and D.~Zenhausern,
  Phys. Lett. B {\bf 671}, 162 (2009);
   H.~Aoki and S.~Iso,
  Phys. Rev. D {\bf 86}, 013001 (2012); 
   Y.~Hamada, H.~Kawai and K.~-y.~Oda,
  {\it ibid.} 
  D {\bf 87}, 053009 (2013);
 M.~Farina, D.~Pappadopulo and A.~Strumia,
  JHEP {\bf 1308}, 022 (2013); 
  M.~Heikinheimo, {\it et al.}, 
  arXiv:1304.7006 [hep-ph];
  G.~F.~Giudice,
  arXiv:1307.7879 [hep-ph];
    G.~Marques Tavares,  M.~Schmaltz and W.~Skiba,
  arXiv:1308.0025 [hep-ph];
    Y.~Kawamura,
  arXiv:1308.5069 [hep-ph].

\bibitem{ccm}
  R.~Hempfling,
  Phys. Lett. B {\bf 379}, 153 (1996);
  K.~A.~Meissner and H.~Nicolai,
  {\it ibid.} 
  B {\bf 648}, 312 (2007);
  W.~F.~Chang, J.~N.~Ng and J.~M.~S.~Wu,
  Phys. Rev. D {\bf 75}, 115016 (2007);
  M.~Holthausen, M.~Lindner and M.~A.~Schmidt,
  {\it ibid.} 
  D {\bf 82}, 055002 (2010);
   L.~Alexander-Nunneley and A.~Pilaftsis,
  JHEP {\bf 1009}, 021 (2010);
 T.~Hur and P.~Ko,
interacting hidden sector,''
  Phys.\ Rev.\ Lett.\  {\bf 106}, 141802 (2011)
  K.~Ishiwata,
  Phys. Lett. B {\bf 710}, 134 (2012);
  I.~Oda,
  Phys. Rev. D {\bf 87}, 065025 (2013);
   C.~Englert, {\it et al.}, 
   JHEP {\bf 1304}, 060 (2013); 
    T.~Hambye and A.~Strumia,
  Phys. Rev. D {\bf 88}, 055022 (2013); 
   C.~D.~Carone and R.~Ramos,
  {\it ibid.} 
  D {\bf 88}, 055020 (2013); 
    A.~Farzinnia, H.~-J.~He and J.~Ren,
  arXiv:1308.0295 [hep-ph];
    E.~Gabrielli, {\it et al.}, 
  arXiv:1309.6632 [hep-ph].

  \bibitem{Iso:2009ss} 
  S.~Iso, N.~Okada and Y.~Orikasa,
  Phys. Lett. B {\bf 676}, 81 (2009);
  Phys. Rev. D {\bf 80}, 115007 (2009).

\bibitem{Iso:2012jn} 
  S.~Iso and Y.~Orikasa,
  PTEP {\bf 2013}, 023B08 (2013).

\bibitem{Chun:2013soa} 
  E.~J.~Chun, S.~Jung and H.~M.~Lee,
  Phys. Lett. B {\bf 725}, 158 (2013).

\bibitem{CW} 
  S.~R.~Coleman and E.~J.~Weinberg,
  Phys. Rev. D {\bf 7}, 1888 (1973).

\bibitem{Pendleton:1980as} 
  B.~Pendleton and G.~G.~Ross,
  Phys. Lett. B {\bf 98}, 291 (1981).

\bibitem{Basso:2010jm} 
  L.~Basso, S.~Moretti and G.~M.~Pruna,
  Phys. Rev. D {\bf 82}, 055018 (2010).

 \bibitem{pdg}
  J.~Beringer et al. (Particle Data Group), 
  Phys. Rev. D86, 010001 (2012). 

\bibitem{Melnikov:2000qh} 
  K.~Melnikov and T.~v.~Ritbergen,
  Phys. Lett. B {\bf 482}, 99 (2000).

\bibitem{Lee:2012xn} 
  H.~-S.~Lee and A.~Soni,
  Phys. Rev. Lett. {\bf 110}, 021802 (2013).

\bibitem{Langacker:2008yv} 
  See, e.g.,  P.~Langacker,
  Rev. Mod. Phys. {\bf 81}, 1199 (2009).

\bibitem{HIO}
  M.Hashimoto, S. Iso and Y. Orikasa, to appear.

\bibitem{flat}
 M.~Holthausen, K.~S.~Lim and M.~Lindner,
  JHEP {\bf 1202}, 037 (2012).

\end{thebibliography}
\end{document}